\documentclass{article}

\usepackage{arxiv}

\usepackage[utf8]{inputenc} % allow utf-8 input
\usepackage[T1]{fontenc}    % use 8-bit T1 fonts
\usepackage{hyperref}       % hyperlinks
\usepackage{url}            % simple URL typesetting
\usepackage{booktabs}       % professional-quality tables
\usepackage{amsfonts}       % blackboard math symbols
\usepackage{nicefrac}       % compact symbols for 1/2, etc.
\usepackage{microtype}      % microtypography
\usepackage{lipsum}
\usepackage{amssymb}
\usepackage{graphicx}
\usepackage{lipsum}  
\usepackage{lineno}  
\usepackage{array}
\usepackage{multirow}
\PassOptionsToPackage{hyphens}{url}
\usepackage{hyperref}

\title{Towards countering hate speech against journalists on social media}
\date{}
\author{
  Polychronis Charitidis\\
  DataScouting\\
  30 Vakchou Street, 54629 \\
  Thessaloniki, Greece \\
  \texttt{pcharitidis@datascouting.com} \\
   \And
  Stavros Doropoulos\\
  DataScouting\\
  30 Vakchou Street, 54629 \\
  Thessaloniki, Greece \\
  \texttt{doro@datascouting.com} \\
     \And
  Stavros Vologiannidis\\
  International Hellenic University\\
  Terma Magnisias, 62124 \\
  Serres, Greece \\
  \texttt{svol@teicm.gr} \\
     \And
  Ioannis Papastergiou\\
  DataScouting\\
  30 Vakchou Street, 54629 \\
  Thessaloniki, Greece \\
  \texttt{ipapaste@datascouting.com} \\
     \And
  Sophia Karakeva\\
  DataScouting\\
  30 Vakchou Street, 54629 \\
  Thessaloniki, Greece \\
  \texttt{soka@datascouting.com} \\
}

\begin{document}
\maketitle

\begin{abstract}
The damaging effects of hate speech on social media are evident during the last few years,  and several organizations, researchers and social media platforms tried to harness them in various ways.  Despite these efforts, social media users are still affected by hate speech.  The problem is even more apparent to social groups that promote public discourse, such as journalists.  In this work, we focus on countering hate speech that is targeted to journalistic social media accounts. To accomplish this, a group of journalists assembled a definition of hate speech, taking into account the journalistic point of view and the types of hate speech that are usually targeted against journalists. We then compile a large pool of tweets referring to journalism-related accounts in multiple languages. In order to annotate the pool of unlabeled tweets according to the definition, we follow a concise annotation strategy that involves active learning annotation stages. The outcome of this paper is a novel, publicly available collection of Twitter datasets in five different languages. Additionally, we experiment with state-of-the-art deep learning architectures for hate speech detection and use our annotated datasets to train and evaluate them.  Finally, we propose an ensemble detection model that outperforms all individual models.
\end{abstract}

\section{Introduction}
\label{Introduction}
Hate Speech is not a new phenomenon. However, before the advent of email, online comments and networking platforms, the threshold to utter it to any effect at all was much higher. People had to draft a letter, buy postage, and send their missive through the mail. Since then, the formulation and dissemination of hate speech have become easy, instant, potentially ubiquitous, public, and therefore much more damaging. In fact, it not only poisons and thus effectively undermines free and open discourse on the Internet, which is bad enough in itself, but also constitutes a threat to the individuals and organizations it is directed at.

% The increasing propagation of hate speech through social media has drawn the attention of governments and researchers. Several papers exist in the literature dealing with the problem of automatic hate speech detection. Hate speech detection methodologies aim to classify any social media post as hate speech or non-hate speech and some even try to derive the type of hate speech.

%%<ADDED>
The increasing propagation of hate speech through social media has drawn the attention of governments and organizations. In May 2016, the European Commission agreed with Facebook, Microsoft, Twitter, and YouTube to a code of conduct \cite{jourova2016code} to prevent and counter the spread of illegal online hate speech. Several other large companies joined the code of conduct later. Although these initiatives mitigate hate speech incidents, the elimination of hate speech needs further work. Following this paradigm, the research community made efforts towards countering hate speech. The latest literature contains an increasing number of works that deal with the problem of automatic hate speech detection. Hate speech detection methodologies aim to classify social media posts into those than contain hate speech and those that do not, while some works even try to identify the type of hate speech.
%%</ADDED>

% In this paper, some of the outcomes of DACHS\footnote{https://hatedetection.com/} (“A Data-driven Approach to Countering Hate Speech”) project will be described. DACHS studies hate speech, focusing on quality journalism as a test case. As professional arbiters of the public sphere, journalists run afoul of hate speech originators practically by default, yet normally without the background or affiliation that usually triggers much of the destructive online communication. Journalists are multipliers of societal discourse, and a side effect of their relative prominence and high audience reach is that they help protect the smaller and weaker actors in the arena of opinions. The main goal of the project is to build an Alert Monitoring Platform that notifies journalists in cases where there is hate speech. The Alert Monitoring Platform allows journalists to monitor hate speech and is posted on social networks. The journalists can configure the threshold over which content with hate speech will be sent to them via mail as aggregated alert reports. Optionally, journalists can contribute to the dataset generation by further annotating such posts. During DACHS, hate speech against journalists in Twitter is studied in 5 languages, English, French, German, Spanish, and Greek. This paper presents ongoing work regarding automated hate speech and personal attack detection as well as the related datasets that were created.

%%<ADDED>
In this paper, we present some of the outcomes of DACHS\footnote{https://hatedetection.com/} (“A Data-driven Approach to Countering Hate Speech”) project. While, all victims of hate speech are equally in need of protection and defense, for the purpose of DACHS, we focus on journalism as a test case. As professional arbiters of the public sphere, journalists run afoul of hate speech originators practically by default. Journalists are multipliers of societal discourse and their relative prominence, and high audience reach makes them vulnerable to hate speech. Report in \cite{unesco} highlights the rapid spread of hate speech against journalists that infringes their freedom of expression. To assist their work and further promote free speech, the DACHS project aims to counter hate speech directed at journalists. One of the main goals of DACHS is to build a Twitter alert monitoring mechanism that notifies journalists about cases where hateful tweets are posted in their Twitter feed. %With this mechanism, journalists are able to monitor their tweets and filter those that contain hate speech. 
Also, they can receive email reports with statistics about hate speech in their timeline at specified time intervals. Optionally, journalists can suggest or flag tweets that they consider to be hate speech. During DACHS, hate speech against journalists in Twitter is studied in 5 languages: English, French, German, Spanish, and Greek.
%%</ADDED>

%This work's contribution is twofold. First, it presents a concise annotation strategy used for the generation of publicly available multilingual hate speech and personal attack Twitter datasets. Second, it uses these datasets to train various state-of-the-art deep learning architectures, while at the same time, proposes an ensemble model and reports the evaluation results. To the best of our knowledge, this is the first effort that creates such a multilingual, diverse and large dataset for hate speech.

%%<ADDED>
This work makes the following contributions. First, it defines hate speech from a journalistic point of view, taking into account examples of hate speech directed at journalists.  The definition is formed after extensive discussions with journalists from the European Journalism Centre\footnote{https://ejc.net/} and its main attributes are that it is simple, concise and accounts for large-scale annotation. Second, it presents a concise two-stage annotation strategy. Both of these stages sample tweets from the collected unlabeled data and generate batches of tweets to be submitted for human annotation. The first stage generates the initial batch using keywords and existing hate speech datasets to filter tweets and annotate them. The second stage is responsible for generating all subsequent batches making use of active learning. 
%strategy that employs an active learning approach. 
This strategy is used to annotate a large pool of journalist-related tweets generating large-scale hate speech datasets in multiple languages. The datasets were made publicly available to assist further research on the field. The third contribution of this work is that it uses these datasets to train various state-of-the-art deep learning models, including an ensemble model that outperforms all individual models. To the best of our knowledge, this is the first work that studies hate speech in multiple languages.
%%</ADDED>

%%<ADDED>
The rest of the paper is structured as follows. Section \ref{Related Work} includes a brief overview of the current state-of-the-art that addresses hate speech. This includes an overview of different hate speech definitions, existing hate speech datasets and automated detection methods. In section \ref{Definition} we present the definition of hate speech that is being used throughout the paper. Sections \ref{Data collection} and \ref{Annotation process}, present the data collection and annotation methodologies respectively. Section \ref{Experimental study} describes the performed experimental study and demonstrates the results of this work. In the last section, we conclude this work and we present some future steps related to hate speech detection.
%%<ADDED>
%%<ADDED>
\begin{table*}[ht]
    \centering
%\scalebox{0.95}{
    \begin{tabular}{|c|c|c|c|c|}
        \hline
         \textbf{Dataset} & \textbf{Source} &  \textbf{Size}   &\textbf{Language} &\textbf{Labels} \\
        \hline
        Wulczyn et al. \cite{wulczyn_ex_2017} &  Wikipedia & 100k & EN &  offensive\\
        \hline  
        Founta et al. \cite{founta2018large} &  Twitter & 80k & EN &  offensive/abusive/hate speech\\
        \hline  
        Davidson et al. \cite{davidson_automated_2017} &  Twitter & 25k & EN & hate speech/offensive\\
        \hline
        Waseem et al. \cite{waseem_hateful_2016} &  Twitter & 16k & EN & sexism/racism\\
        \hline 
        Sharma et al. \cite{sharma_degree_2018} &  Twitter & 9k & EN & multiple hate speech classes\\
        \hline        
        Gibert et al. \cite{gibert_hate_2018} &  Other & 10k & EN & hate speech\\
        \hline       
        ElSherief et al. \cite{elsherief_peer_2018} &  Twitter & 28k & EN & multiple hate speech classes\\
        \hline        
        Kwok et al. \cite{kwok_locate_2013} &  Twitter & 24k & EN & racism\\
        \hline  
        Ross et al. \cite{ross_measuring_2017} &  Twitter & 541 & DE & racism\\
        \hline  
        Wiegand et al. \cite{wiegand2018overview} &  Twitter & 9.5k & DE & insult/abuse/profanity\\
        \hline  
        Del Vigna et al. \cite{del_vigna12_hate_2017} &  Facebook & 17.5k & IT &  multiple hate speech classes\\
        \hline
    \end{tabular}
%    }
    \caption{Related dataset information}
    \label{tab:related_datasets}
\end{table*}
%%</ADDED>

\section{Related Work}
\label{Related Work}
\subsection{Definitions}

% One of the challenges in studying negative online behavior and hate speech in particular, is the lack of a clear, common definition \cite{saleem_web_2017}. In recent literature, many authors studied cyberbullying \cite{hosseinmardi_analyzing_2015,zhong_content-driven_2016}. Work in \cite{wulczyn_ex_2017} employs the term personal attack to describe offensive online behavior, while other studies focus on offensive or abusive speech and online harassment \cite{park_one-step_2017,davidson_automated_2017,nobata_abusive_2016}. The actual term hate speech is used in many previous works \cite{warner_detecting_2012,burnap_cyber_2015,djuric_hate_2015,kwok_locate_2013,sharma_degree_2018,waseem_hateful_2016,gibert_hate_2018}. Even though these definitions share many common characteristics, in many cases, there are distinct differences even between definitions that use the same term to describe negative online behavior.
%%<ADDED>
 One of the challenges in studying negative online behavior, and hate speech in particular, is the lack of a clear, common definition \cite{saleem_web_2017}. Generally speaking, hate speech could be described as the expression of hatred towards an individual or group of individuals because of a characteristic they share, or a group to which they belong. In \cite{wulczyn_ex_2017} the term personal attack is used to describe offensive online behavior, while other studies focus on offensive or abusive speech and online harassment \cite{park_one-step_2017,davidson_automated_2017,nobata_abusive_2016}. Other works like \cite{jane2014your}, address particular types of online harassment or hate, like misogyny. The actual term hate speech is used in many previous works \cite{warner_detecting_2012,burnap_cyber_2015,djuric_hate_2015,kwok_locate_2013,sharma_degree_2018,waseem_hateful_2016,gibert_hate_2018}. Even though these definitions share many common characteristics, there are distinct differences even between definitions that are using the same term to describe negative online behavior. Authors in \cite{assimakopoulos2017online} argue that even more formal definitions of illegal hate speech, like the EU definition \cite{eudef} or the United Nations definition \cite{ohchr}, contain words that are open to interpretation. Illegal hate speech is further examined in another European project\footnote{http://mandola-project.eu/} that takes into account the heterogeneity and complexity of different legislations. Furthermore, authors in \cite{citron2014hate} thoroughly discuss and define hate crimes in cyberspace.
 
 In our case, our intention is to find and work with a notion of hate speech that takes into consideration the journalistic point of view and at the same time is easy to understand. To this end, we examined the related literature and created a new definition that is in line with the project's requirements.
 %%</ADDED>
 
\subsection{Datasets}
With the advent of social media, research on hate speech was intensified during the last few years. A critical step to achieve further progress in
the detection of online hate speech is the availability of large scale
datasets. There have been relatively few efforts focusing on the creation of
hate speech datasets from social media. Davidson et al. \cite{davidson_automated_2017}
collected Twitter data using a hate speech lexicon compiled with the help of Hatebase.org
in English. They employed crowd-sourcing to label tweets into three
categories: hate speech, offensive language, and those with neither.
Waseem et al. \cite{waseem_are_2016,waseem_hateful_2016} provide
a hate speech dataset, which contains 16k tweets, and describe the respective annotation procedure, in which
an initial manual search was conducted on Twitter to collect
common slurs and terms about religion, sexual orientation, gender, and
ethnic minorities. The dataset was then manually annotated regarding the existence of sexism or racism. Sharma et al. \cite{sharma_degree_2018}
collected a set of 9k tweets containing harmful speech and they manually
annotated them in three classes based on their degree of hateful intent.
The authors of \cite{gibert_hate_2018} crawled data from a white supremacy
forum to extract and to manually annotate over 10k sentences
as hate speech or not. Authors in \cite{elsherief_peer_2018} describe
a multi-step classification process and they provide a comprehensive hate
speech dataset containing more than 28k tweets with various
types of hate related to sexual orientation, gender, ethnicity, etc.
Finally, in older works, many researchers have relied on creating their own
hand-coded hate speech datasets as in  \cite{kwok_locate_2013,warner_detecting_2012}.

The majority of hate speech related studies focus on the English
language. However, in \cite{ross_measuring_2017,wiegand2018overview}, hate speech against refugees is studied in the German language. The authors in \cite{del_vigna12_hate_2017}
crawled Facebook comments from public Italian pages and annotated them with a variety of hate categories to distinguish different notions of hate speech.
 
In addition, there are a lot of datasets addressing offensive and
toxic online behavior. Kaggle's Toxic Comment Classification Challenge
dataset \cite{noauthor_toxic_nodate} consists of 150k Wikipedia
comments annotated for toxic behavior. Kaggle hosts additional large
scale toxic speech datasets like \cite{noauthor_quora_nodate,noauthor_detecting_nodate}.
Studies in \cite{wulczyn_ex_2017,founta2018large} use crowdsourcing
to provide abuse-related annotation on 100k English Wikipedia comments
and 80k tweets respectively. Finally, smaller datasets as in \cite{nobata_abusive_2016,burnap_cyber_2015}, focus on the annotation of toxic versus non-toxic online comments.

%%<ADDED>
Table \ref{tab:related_datasets} presents a summary on the datasets available in related literature.
%%<ADDED>

\subsection{Detection methods}

Existing hate speech detection methods address the problem as a supervised
classification task \cite{schmidt_survey_2017}. Traditional methods
rely on manually designing and encoding features of textual data into
feature vectors, that are used as inputs to algorithms, such as Naive
Bayes, Logistic Regression, SVM and Random Forest. These methods are
adopted by numerous hate and offensive speech detection studies, such as  \cite{burnap_cyber_2015,davidson_automated_2017,djuric_hate_2015,kwok_locate_2013,mehdad_characters_2016,warner_detecting_2012,waseem_are_2016,waseem_hateful_2016,xiang_detecting_2012,gao_detecting_2017}.
These studies experiment with various features, including bag-of-words
representation, character-level, word-level n-gram features, syntactic
features, linguistic features, and comment embedding features.

Following the more recent deep-learning paradigm, several studies
use neural networks to detect hateful and toxic content. Neural networks
learn abstract feature representations from input data through multiple
stacked layers. The key difference from traditional models is that deep learning models automate the feature extraction process and the multi-layer structure provides more efficient feature representations.
Many studies have shown that deep learning and neural network methods
outperform traditional methods on hate speech detection tasks \cite{gamback_using_2017,park_one-step_2017}.
The most popular network architectures are Convolutional Neural Networks
(CNN) and Recurrent Neural Networks (RNN). In the context of hate speech
classification, CNN extracts meaningful features from word or character
combinations \cite{badjatiya_deep_2017,gamback_using_2017,park_one-step_2017,zhang_hate_2018},
while RNN learns word or character dependencies in sequences of words
\cite{gao_detecting_2017,badjatiya_deep_2017,del_vigna12_hate_2017,founta2019unified}.
Combinations of CNN and RNN models are also applied in \cite{zhang_detecting_2018}.

\newcolumntype{C}[1]{>{\centering\arraybackslash}p{#1}}
\begin{table*}[ht]
    \centering
    \begin{tabular}{|C{13.7cm}|C{1.3cm}|}
        \hline
         \textbf{Tweet} &   \textbf{Type} \\
        \hline
        @USER They shud DEPORT that mot******ker Back to Iran &  1\\
        \hline
        @USER Please somebody, kill him but before you do torture him to death! & 2\\
        \hline
        @USER Whats the matter you cowards have someone disagree with you and your coward journalists and boom they are taken off platforms for daring to have an opinion against you.... I hope you all rot in hell Wall Street Journal for the cowards you are! &1\\
        \hline
        @USER Kill the NBC journalists !!! & 2\\
        \hline
        @USER Let me fix this for you.  Chicken good...Queers Bad.  Solved it for you. & 3\\
        \hline
        @USER @USER ACCORDING TO WHITE PEOPLE...		When whites kill = Lone Wolf Mental Illness (Even though they've killed all over the globe for Centuries).		When black people kill = Entire black race is violent . & 4\\
        \hline
        @USER @USER Just cut the diplomatic ties to UK. We Germans call them island monkeys. Or island apes. &4\\
        \hline
        
    \end{tabular}
    \caption{Hate speech tweets and the type of hateful attack that corresponds to the second bullet of the definition in Section \ref{Definition}}
    \label{tab:english_examples}
\end{table*}

\section{Definition}
\label{Definition}

To consistently annotate a large Twitter corpus, there is a need for a clear and simple definition. We define hate speech in a way that is easy for annotators to label tweets but also for other non-expert groups, like journalists, to further enhance the dataset or provide feedback.
% Need to add more details to this
The proposed definition is formed after extensive discussions with journalists from the European Journalism Centre through a small focus group and continuous evaluation  and feedback from journalists.
After looking at hundreds of hateful tweets and several meetings, it was decided that the presence of hate speech should be concluded by answering to two simple key questions. These questions refer to the tweet content and they are presented below:
\begin{itemize}
\item Does it target a person or group?
\item Does it contain a hateful attack?
\begin{enumerate}
\item Violent speech
\item Support for death/disease/harm
\item Statement of inferiority relating to a group they identify with (like
LGBTQI)
\item Call for segregation
\end{enumerate}
\end{itemize}
A positive answer to both bullets should make the annotator flag the post as hate speech. The second question can refer to any of the four subcategories that are listed above. This definition was evaluated with a larger group of journalists and proved to be concise and easy to understand. 

To give a better intuition about the hate speech definition, we provide some examples of annotated tweets in English language. The examples are listed in Table \ref{tab:english_examples}. Note that all of these tweets comply with the first requirement of the above definition, meaning that these tweets target a person or a group. Table \ref{tab:english_examples} also identifies the type of hateful attack on each tweet in four classes, as described in the second bullet of the definition.

% The following definition from \cite{wulczyn_ex_2017} is used for identifying personal attack:
% \begin{itemize}
% \item Does it contain a personal attack or harassment?
% \begin{itemize}
% \item Targeted at the recipient of the message (i.e. you suck).
% \item Targeted at a third party (i.e. Bod sucks).
% \item Being reported or quoted (i.e. Bod said Henri sucks).
% \item Another kind of attack or harassment.
% \end{itemize}
% \end{itemize}

\section{Data collection}
\label{Data collection}

% The main aim of this work is to create datasets of hate speech in a journalistic context and in English, French, German, Greek and Spanish. The first step of the data collection process consists of gathering a list of journalism-related Twitter accounts for each language, that are used as sources for tweet retrieval. A straightforward way to compose such a list was to manually identify a list of well-known accounts of journalists and news outlets and focus data collection on those accounts. However, preliminary experiments showed that the volume of data that could be collected following this approach is limited, at least by using the standard version of Twitter’s Search API\footnote{https://developer.twitter.com/en/docs/tweets/search/api-reference/get-search-tweets.html} which does not provide historical data. 

%%<ADDED>
One of the goals of this work is to create multilingual hate speech datasets consisting of tweets that originate from a journalistic context, alongside with binary annotations about hate speech. To generate these datasets, a large list of unlabeled tweets is collected. As a side-note, during data collection and management all EU General Data Protection Regulations were followed.

The data collection process started with the creation of a list of journalism-related Twitter accounts. In a subsequent step, we retrieve the tweets related to these accounts to assemble the pool of unlabeled data. Apart from English, which is the most common language in the related literature, this process is applied for French, German, Greek and Spanish. 

 A straightforward way to compose such a list is to manually identify a list of well-known accounts of journalists and news outlets and focus the data collection on those accounts. However, preliminary experiments showed that the volume of data that could be collected following this approach is limited, at least by using the standard version of Search API\footnote{https://developer.twitter.com/en/docs/tweets/search/api-reference/get-search-tweets.html} that does not provide historical data.
%%<ADDED>

To overcome this issue we identify a larger number of journalist-related accounts by using Twitter lists.  Twitter lists\footnote{https://help.twitter.com/en/using-twitter/twitter-lists}
are curated groups of Twitter accounts and are usually centered around
specific themes. In order to find Twitter lists related to journalism,
we collected the Twitter accounts of well-known news outlets and journalism-related
organizations and automated the process of fetching all list members. Despite the fact that the majority of the collected
accounts are related to either journalists or news outlets, we also noticed
the presence of non-journalistic accounts, which could potentially tamper
with the journalistic focus of our data collection. To tackle this issue,
we manually filtered the irrelevant accounts. Since annotating
(for relevance with journalism) the full list of accounts would involve
a considerable manual effort, we prioritized the annotation of
the most popular accounts, since those accounts usually attract a
larger number of tweets. In the first iteration of account collection,
we annotated the accounts using the following classes: a) journalist,
b) news outlet, c) irrelevant. The initial seed consists of 200 manually
collected journalism-related accounts for each language. Table \ref{tab:retrieval_stats} lists the number of journalism-related accounts per language.

Having a validated list of journalism-related Twitter accounts for each language, we set up, as the next step, a mechanism to collect tweets from the feeds of these accounts. To this end, the Twitter Search API is used, which returns a sample of Tweets posted in the past 7 days. 
The API is rate-limited at 180 requests per
15-min window when user authentication is used and at 450 requests
per 15-min window when application authentication is used. We used application authentication having also in mind that each call  can return a maximum of 100 tweets.
Despite these limitations, by effectively utilizing the API within the imposed call rate limitations, we manage to collect a large number of tweets that is sufficient for creating a sizeable, journalism-oriented hate speech database. Typically, the Twitter Search API query consists of a series of keywords that should be contained in the set of returned tweets, along with account-based search operators. We opted for using search queries that do not restrict the tweet contents and only used account-based search operators to limit results. More specifically, we used the \textit{to:} (e.g., to:BBCNews) and the “@” (e.g. @BBCNews) operators in order to collect tweets authored in reply to and tweets mentioning those specific accounts. 

Specifically, every 15 minutes, the module sequentially performed $N=450-N_{safe}$ API requests, evenly distributed across the 15-minute window. $N_{safe}[0,450)$ corresponds to a safety parameter that is used to avoid pushing the API to its limits. We use $N_{safe}=200$ and thus a maximum of $N=250$ calls is performed per every 15-min window. Since each call is associated with one account, and the number of calls that can be performed within each 15-min window is much smaller than the pool of target accounts, an account prioritization/selection mechanism is implemented. 

%A naive approach would consist of sequentially iterating over the list of accounts in batches of N accounts per time window and fetching the latest 100 tweets that each account has received using the $count=100$ and the $result\_type=recent$ call parameters. This approach would ensure that content from all accounts would be regularly fetched in even time intervals but had the disadvantage that it is rate-agnostic (i.e., ignored the fact that each account could receive tweets at a different rate) and would, therefore, result in performing many requests that return much fewer tweets than the maximum of 100 tweets per request. A better approach would be to limit the pool of target accounts to the K most followed ones, under the assumption that the number of followers is indicative of the rate at which an account receives new tweets. Although this approach would improve the effectiveness of the calls, it would have two disadvantages: a) it would limit the data collection to only K instead of N accounts and b) it would assume a static incoming tweet rate for each account while the rate is highly dynamic.

This prioritization approach fetched data from all accounts and measured an estimate of the rate of incoming tweets. Using this estimate, the available API calls are distributed on accounts, which are expected to have received a sufficient number of new tweets since the last time that they were queried. Specifically, whenever a new API call is performed, an estimated number of available new tweets is calculated for each account based on the account’s estimated incoming tweet rate and the time that has passed since the last time it was fetched. Then, one account is randomly selected among those whose estimated number of available new tweets is larger than a user-specified threshold (for which reasonable values should lie close to 100, the maximum number of results per request). After a call is performed, the incoming tweet rate for each account is updated by dividing the number of returned tweets with the time difference between the newest and the oldest tweet (a very low rate is assigned in case a call returns zero results). Moreover, the last fetched timestamp is recorded to facilitate the calculation of the estimated number of incoming tweets. Initially, all accounts are assigned a fixed, high incoming tweet rate to ensure that all accounts will be fetched at least once. This approach ensured that all accounts were queried proportionally to their actual, dynamic incoming tweet rate, thus maximizing the amount of tweets that can be collected.

We apply this approach for every language, mining tweets from the corresponding pool of journalistic Twitter accounts and store the retrieved tweets in a mongoDB database. To ensure that tweet language matches the language of the query accounts, we inspect the tweet metadata information.

\begin{table}
    \centering
    %    \scalebox{0.93}{
    \begin{tabular}{|c|c|c|}
        \hline 
        \textbf{Language} & \textbf{Accounts} & \textbf{Tweets}\\
        \hline 
        \hline 
        EN & 12,306 & 92,324,248\\
        \hline 
        DE & 3,436 & 12,436,132\\
        \hline 
        ES & 1,765 & 49,453,601\\
        \hline 
        FR & 2,794 & 34,118,951\\
        \hline 
        GR & 3,577 & 2,147,668\\
        \hline 
    \end{tabular}
    \caption{\label{tab:retrieval_stats}Total number of journalism-related Twitter accounts and total number of tweets retrieved per language}
\end{table}

The total number of the collected unlabeled tweets per language is listed in Table \ref{tab:retrieval_stats}. We denote the unlabeled pool of tweets by \(\mathcal{U}\).
The unlabeled pool of tweets for each language is denoted by  $U_i$ where $i \in \{EN, DE, ES, FR, GR\}$.
Notice that there are significant differences in the number of tweets retrieved per language. This is expected due to the different number of query accounts used to retrieve tweets, the language popularity and the Twitter usage per country.  The total collection period was approximately 6 months (1/10/2018 - 8/5/2019).

\section{Annotation process}
\label{Annotation process}
In this Section, we describe the annotation process for labelling the pool of tweets \(\mathcal{U}\). The purpose of this process is to create a labeled dataset denoted by \(\mathcal{L}\). Note that \(\mathcal{L}\) consists of 5 different language datasets denoted by $L_i^k$ where $i \in \{EN, DE, ES, FR, GR\}$ and $k$ denotes the current number of annotated batches that is included in the dataset. 

The annotation process consists of two sampling stages, the initial sampling stage and the active learning sampling stage. In  the initial sampling stage, we use two approaches described in the following subsection, and select a defined number of tweets from the $U_i$ pool to generate the initial annotation batch for each language $i$. We denote annotation batches for each language by $B_i^k$ where $k$ is the number of the generated batch. For the initial sampling stage $k=1$. Each $B_i^1$ in this stage, is submitted for human annotation. After, the annotation of the initial $B_i^1$, the first labeled datasets $L_i^1$ are generated.

The active learning sampling stage,generates all the subsequent $B_i^k$ with $k>1$, for further annotation and dataset expansion, through an iterative process. In the first iteration, we use $L_i^1$ generated from the initial sampling stage, and we employ an active learning sampling approach to generate $B_i^2$ from $U_i$. We then submit  $B_i^2$ for annotation and we expand the $L_i^1$ to $L_i^2$ by appending the new annotated batch. We repeat this process until there are no more tweets available in the $U_i$ or we exhaust the annotation budget.

%In the next subsections, we explain our comprehensive process that ensures high-quality annotations but also accounts for large-scale dataset generation.

\subsection{Initial sampling stage}
\label{Initial annotation batches}

% The annotation was performed in batches of tweets. Namely, the annotators responsible for the respective languages were given a relative small number of tweets for annotation. After conducting a brief preliminary annotation round in various languages using random tweets from the journalist-related pool, we observed an apparent scarcity of hateful content among annotated tweets. Table \ref{tab:hate_speech_percent} shows that there are only 8 tweets that annotated as hate speech in a total of 2000 annotated tweets that were randomly retrieved from the English database. Based on this observation, it is evident that we needed to develop a process, which would generate annotation batches with a higher percent of tweets that express hate.

By observing Table \ref{tab:retrieval_stats}, we notice that there is a large number of unlabeled tweets in each language. To submit such sizable data for human annotation, is not only overwhelming to the annotators, but makes the whole process very expensive and hard to supervise. Additionally, we expect that the number of positive annotations will be very small compared to the negative annotations. To validate this, we conduct a brief preliminary annotation round, annotating 2000 random tweets per language and inspecting the number of positive annotations. As expected, we observe an apparent scarcity of positive annotations. Table \ref{tab:hate_speech_percent} shows that for the preliminary batch in English language, among 2000 annotations, only 0.4\% are positive. Based on these observations, it is evident that we need to develop a process, which would generate batches for annotation of manageable size and exhibiting a higher positive annotation ratio.

To mitigate the lack of positive annotations, we consider two different approaches for generating the annotation batches of tweets ($B_i^k$). The first and the most popular approach in the literature, is using keyword-based sampling and thus creating the annotation batch with tweets containing specific keywords. In the hate speech context, these keywords are usually offensive words or words that can be used to express hatred. The second approach, which is novel in the literature, is to use existing hate speech datasets in order to train hate speech detection models and, in a subsequent step, apply these models to the pool of unlabeled tweets and sample the tweets with higher hate speech probability. For this task we train a CNN model, which is described later in section \ref{Models}. In a sense, this approach shares similarities with transfer learning \cite{torrey2010transfer}, where we essentially transfer the established hate speech definition from another work to sample our data. We refer to this approach as dataset-based sampling. One evident shortcoming of this approach is that it requires the availability of such datasets, something that is possible only for a few languages.

Both of these approaches introduce bias in the sampled tweets \cite{wiegand2019detection} and consequently in the generated datasets. %A recent work  argues that the most popular hate speech datasets suffer from topic bias. 
It is obvious that dataset generation with such extreme imbalance between classes, as the one presented in this work, requires a sampling approach that is inclined towards the minority class. %This requirement is even more apparent in cases where the annotation budget is limited. 

Although bias is inevitable, there are ways that it can be moderated. To moderate bias for the keyword-based approach, we compile a large list of keywords. These lists included offensive slang, phrases and words that could potentially express hate when used in the appropriate context and keywords related to several forms of possible discrimination (religion, gender, refugees etc). For the dataset-based approach, we adopt three different approaches to tackle bias. First, we use datasets from works that have less strict definition of hate speech compared to ours. Second, in cases where the dataset-based approach is not applicable due to lack of existing datasets we combine multiple different datasets with diverse definitions and scopes. Third, we use a loose classification threshold in the resulting CNN model, sampling tweets from a wide range of hate speech probability. Finally, to further reduce bias in the generated annotation batch, we concatenate random sampled tweets for both keyword-based and dataset-based approaches.

We apply the dataset-based sampling approach to generate $B_{EN}^1$ and $B_{DE}^1$ by using some of the related datasets in Table \ref{tab:related_datasets}.
%To generate the initial $B_i^1$, we apply the dataset-based sampling for English and German language. There are multiple available datasets for these languages as it is demonstrated in Table \ref{tab:related_datasets}.
We apply the keyword-based sampling for the Spanish, French and Greek due to lack of related datasets. %The reason we apply the keyword-based approach is because there are no available datasets in the corresponding languages. On the other hand, such hate speech datasets exist for English and German language as shown in Table \ref{tab:related_datasets}.

Specifically, for the English language, we train a CNN model using the dataset presented in \cite{davidson_automated_2017}. Although their definition of hate speech differs from the one described in this work, the dataset is suitable for training, since \cite{davidson_automated_2017} establishes a loose definition of hate speech compared to ours (i.e  a lot of offensive and insulting tweets are considered as hate speech). For the case of the German language, two different datasets are combined. The first dataset  \cite{ross_measuring_2017}, included tweets referring to refugees and included binary hate speech annotations. 
%Most of the hate speech in this dataset is about racism.
Additionally, the GermEval 2018 dataset \cite{wiegand2018overview} is used, which is a series of shared task evaluation campaigns that focused on natural language processing for the German language. Just like the English case, a CNN model is trained using this dataset. To generate the first annotation batch for each one of these languages,  we applied the corresponding CNN model to the pool of unlabeled tweets calculating the hate speech probability for each unlabeled tweet. Then we randomly sample 8000 tweets with hate speech probability that falls in range [0.2-1.0] for each language. This wide range is chosen to reduce the bias of the generated batch as discussed previously.

For the rest of the languages (Spanish, French, Greek), we employ the keyword-based approach. For each language we assemble 
%a list of keywords. A small list of such keywords can introduce high bias in the resulting dataset, because they can potentially capture specific types of hate speech. To mitigate bias,
a large list of keywords, including 500 to 1000 keywords per language. These lists included offensive slang, phrases and words that could potentially express hate when used in the appropriate context and keywords  related to several kinds of possible discrimination (religion, gender, refugees etc). We use these keywords to sample from the pool of unlabeled tweets and fetch up to 8000 tweets per language.

% To generate the first annotation batch, the following process was used (for each language) with a goal to increase the probability of retrieving a tweet with hate speech in comparison with random selection. If a hate speech detection model could be obtained by using available datasets for the each language, we applied it to the corresponding pool of unlabeled tweets calculating a hate speech probability for each unlabeled tweet. Then 8000 tweets were randomly retrieved corresponding to a wide range of hate speech probability [0.2-1.0]. This wide range was chosen to reduce the bias of the generated batch. Note that for the case of English we had 2 available models, so we applied them both to the pool of unlabeled tweets and we retrieved those which satisfy the hate speech probability constraint for both models. If a hate speech detection model could not be created, a keyword list as described in the previous paragraph was used to fetch up to 8000 tweets.

For each language, 2000 additional randomly selected tweets are concatenated to the corresponding batch, in order to further mitigate bias imposed by the keyword-based and dataset-based approaches. This leads to the generation of the initial $B_i^1$ that contain 10000 tweets per language. Note that tweets in $B_i^1$  are removed from $U_i$, in order not be retrieved again during the creation of the next batches.

After the generation of initial batches, they are submitted to the corresponding annotators to produce $L_i^1$. We describe the manual annotation process in the next subsection. Table \ref{tab:hate_speech_percent} shows that annotated $B_i^1$ exhibits a significant increase in the ratio of positive annotations compared to the preliminary annotation batch. This is expected because keyword-based and dataset-based sampling favour the hate speech class.

\begin{table}[ht]
    \centering
    %\resizebox{\columnwidth}{!}{
    \begin{tabular}{|c||c|c|c|c|}
    
        \hline 
        \multirow{2}{*}{\textbf{Language i}} & \multicolumn{3}{c|}{\textbf{Batch type}} \tabularnewline
        \cline{2-4} 
        & \textbf{Preliminary} & \textbf{$B_i^1$} & \textbf{$B_i^2$} \tabularnewline
        \hline
        \hline 
        EN & 0.4\% & 1.9\% & 6.33\%\tabularnewline
        \cline{2-4} 
        \hline
        DE & 0.2\% & 1.13\% & 3.47\%\tabularnewline
        \cline{2-4} 
        \hline
        ES  & 0.12\% & 0.89\% & 2.51\%\tabularnewline
        \cline{2-4} 
        \hline
        FR  & 0.3\% & 2.36\% &7.37\%\tabularnewline
        \cline{2-4} 
        \hline
        GR   & 0.13\% & 0.70\% & 1.32\%\tabularnewline
        \cline{2-4} 
        \hline

    \end{tabular}
    
  %  }
    \caption{\label{tab:hate_speech_percent} Ratio of positive annotations in different batches and languages. Preliminary batch includes 2k randomly sampled tweets for each language. $B_i^1$ include 10k tweets selected with keyword-based or dataset-based approach in the initial sampling stage. $B_i^2$ include 10k tweets from the first batch of active learning sampling stage}
\end{table}

\subsection{Manual annotation}
\label{Annotation description}

After generating each batch of tweets for each language, we proceed to the manual annotation task. This process is described in this subsection.

There are many annotation methodologies that can be used on a large corpus of data in the related
literature. For this work, we follow the findings reported in \cite{khetan2017learning}, which claims that it is better to allocate the annotation budget to label as many examples as possible when the annotation quality is above a specified threshold. Based on this, we perform the annotation task submitting one annotation per tweet and thus, utilizing the annotation budget in order to annotate as many tweets as possible. For each language we used experienced annotators, proficient in the corresponding language, familiar with social media and acquainted with the colloquial nature of online conversations. Additionally, we follow a strict quality assurance methodology assigning a supervisor role in the process. The supervisor's job is to validate the annotations following a quality control methodology. In total five annotators (one for each language) and one supervisor are employed for the annotation task.

 The supervisor and annotators of each language perform preliminary annotations to make sure that the definition of hate speech are correctly understood. This process also includes discussions about the particularities of each language. For each language, the corresponding annotator is asked to flag hate speech in tweets with a yes or no answer. In cases where the annotator is unsure about the answer, he/she could flag the tweet, and on a second stage, this post would be discussed with the supervisor. 

To ensure the quality of the annotation, we follow the quality control methodology as described in ISO 2859 and ANSI/ASQ Z1.4-2003. We used level II, i.e., the normal severity level, and thus a lot of 1,000 annotations is considered of acceptable quality if the error rate does not exceed 4\%. To determine this, a single sampling size of 80 annotations out of 1,000 tweets is used. Whenever an annotator completes 1,000 annotations, the supervisor of the process evaluates 80 random samples of them, and if more than 7 annotations are erroneous,  the whole lot is rejected, and careful instructions are given to the annotator. The annotator will then annotate anew the tweets.

\subsection{Active learning sampling stage}
\label{Active learning}

To generate the initial batch of tweets for each language, we use the keyword-based and the dataset-based sampling approaches, and then human annotators label each language batch, as described in previous Subsections. At this point, $L_i^1$ contain 10k labeled tweets for each language. We use the current annotated sets and train new detection models in order to generate new $B_i^2$. This approach is similar to the dataset-based sampling approach, although there are some key differences. Detection models are created for all examined languages (and not just for English and German) with our own annotated datasets. At the same time, these detection models will detect hate speech that matches the definition that is presented in Section \ref{Definition} and not ones that are transferred from other works.

Although this is a valid approach that can be efficiently used to sample tweets from the unlabeled pool \( \mathcal{L} \), it does not guarantee that the sampled tweets will add additional value to the models. For instance, if a hate speech classifier is highly confident about a tweet, then this tweet will not contribute to the learning process and might even hurt generalization performance. Also, this implies that this specific tweet is very similar with other tweets existing in the dataset that were used to generate the classifier. Motivated by this assumption, an active learning mechanism is adopted to generate annotation batches of tweets that would essentially contribute to the ability of the models to learn.

Pool-based active learning \cite{lewis1994sequential} relies on an initial small set of labeled instances \( \mathcal{L} \), and a larger set of unlabeled ones \( \mathcal{U}\). Batches of informative training samples are iteratively selected from \( \mathcal{U} \) and added to \( \mathcal{L} \), with respect to some selection mechanism, after a query about their actual label to an annotator. This approach is motivated in many modern machine learning applications, where unlabeled data may be abundant, but labels are difficult or expensive to obtain.

\begin{figure*}[t]
  \centering
  \includegraphics[scale=0.46]{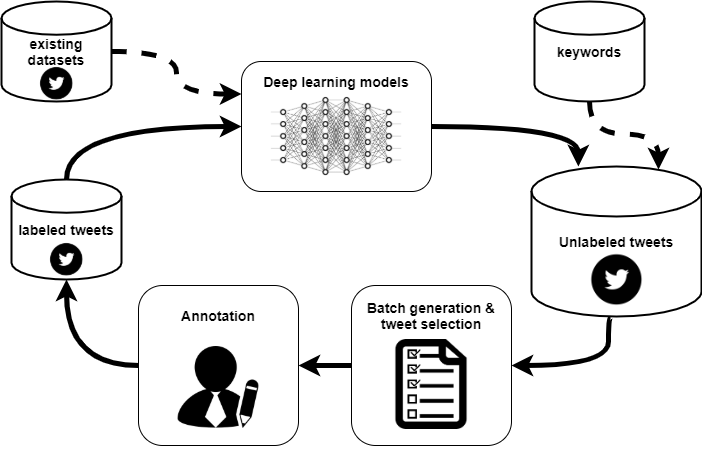}
  \caption{A schematic representation of the annotation process (dotted arrows used only for the initial sampling stage)}
  \label{fig:methodology}
\end{figure*}

Initially, uncertainty sampling \cite{lewis1994sequential} is considered as a  selection mechanism. In this setup, an active learner 
%queried the tweets about which it is the most uncertain of how to label. This approach is very straightforward for probabilistic learning models. In our binary classification model, uncertainty sampling 
selects tweets whose posterior probability is near 0.5. This approach proved problematic for our scenario, since hate speech datasets are imbalanced, and thus the probability distributions are skewed towards the dominant class. Having this in mind, and also the fact that multiple deep learning models would be trained for comparison and evaluation, the best approach proved to be the query-by-committee \cite{seung1992query} algorithm. The query-by-committee approach involves
maintaining a committee \( \mathcal{C} = \{\theta ^{(1)},...,\theta ^{(c)}\}\) of models, which are all trained on
the current labeled set \( \mathcal{L} \), but represent competing hypotheses. This is applicable to our scenario since each deep learning model architecture captures different semantic and syntactic components of the tweet, even if they are trained using the same dataset.
Then, each committee member is allowed to vote on the labels of query candidates. The most informative tweet to be sent to annotators is considered to be the one that committee members disagree upon the most.
The average Kullback-Leibler (KL) divergence \cite{mccallumzy1998employing}:
\[x_{KL}^{*} = argmax\frac{1}{C}\sum_{c=1}^{C}D(P_{\theta^{(c)}}||P_{C})),\]
where:
\[D(P_{\theta^{(c)}}||P_{C}))= \sum_{i}P_{\theta^{(c)}}(y_{i}|x)\log \frac{P_{\theta^{(c)}}(y_{i}|x)}{P_{C}(y_{i}|x)}\]
is used for measuring the level of disagreement among the classifiers.
Here \(\theta^{(c)}\) represents a particular model in the committee, and \( \mathcal{C}\) represents the whole committee, thus \( P_{C}(y_{i}|x)= \frac{1}{C}\sum_{c=1}^{C}P_{\theta^{(c)}}(y_{i}|x)\) is the “consensus” probability that \(y_{i}\) is the correct label. KL divergence \cite{kullback1951information} is
an information-theoretic measure of the difference between two probability distributions. So, this disagreement measure considers the most informative query to be the one with the largest average difference between the label distributions of any of the committee members and the consensus.

\newcolumntype{P}[1]{>{\centering\arraybackslash}p{#1}}

\begin{table}
    \centering
    %\resizebox{\columnwidth}{!}{
    %\scalebox{0.93}{
        \begin{tabular}{|c|P{3.6cm}|c|c|c|}
            \hline 
            \multicolumn{2}{|c|}{\textbf{Train dataset}} & \multirow{2}{*}{\textbf{Test dataset}} &
            \multirow{2}{*}{\textbf{macro-F1}}\\
            \cline{1-2}
            \textbf{Initial} & \textbf{Additional} &    & \\ 
            \hline 
            \hline 
            8000 & -  & 2000   & 0.43\\
            \hline 
            8000 & 2000 (random) & 2000    & 0.43\\
            \hline 
            8000 & 2000 (hate probability)  & 2000    & 0.45\\
            \hline 
            8000 & 2000 (active learning) & 2000    & 0.47\\
            \hline 
        \end{tabular}
    %}
    \caption{\label{tab:active_learning_exp} Different evaluation experiments in English language. We evaluate the models created with an initial train set and additional sets generated with different techniques.}
\end{table}

To apply this process to our work, for $k$-th iteration we train various deep learning models for hate speech detection in each language, using the annotated tweets $L_i^{k-1}$ that are generated from the previous iteration $k-1$. We denote the committee of models for each language $i$ by $C_i$.
Specifically, the active learning sampling stage is described in the following bullets and the entire annotation process is presented in Figure \ref{fig:methodology}.

Thus, for each language $i$:
\begin{enumerate}
\item  We train the models described in Section \ref{Models} and generate the committee of classifiers $C_i$, using the previous batch of annotated tweets $L_i^{k-1}$. Note that in this step, we do not include the ensemble model to the committee $C_i$.
\item For every tweet in the unlabeled pool $U_i$, we calculate each model's \(\theta^{(c_i)}\) hate speech probability and compute the average KL divergence.
 \item For every tweet in the unlabeled pool $U_i$, we calculate the ensemble model's hate speech probability.
\item We sample 8000 tweets from $U_i$ that have the highest KL divergence and the ensemble output probability is higher than 0.2. We also randomly sample 2000 tweets to generate the batch $B_i^k$ with a total of 10000 tweets. Note that the sampled tweets are removed from $U_i$. Also, in the final iteration the size of the batch might be smaller than 10000 in case where there are no available tweets left in $U_i$.
\item We submit $B_i^k$ for annotation and we append the annotated batch to the pool of labeled tweets  $L_i^k$. 
\item The process is repeated from step 1, using  $L_i^k$  to retrain the classifiers and generate the next batch for annotation. We stop this process if there no available tweets in $U_i$ or the annotation budget is exhausted.
\end{enumerate}

In Table \ref{tab:hate_speech_percent}, it is evident that the first annotation batch created with the active learning approach ($B_i^2$) had a higher percent of hate speech compared to the initial or the preliminary batch. This is expected because in the active learning setup, we use models that are trained on datasets annotated using our definition of hate speech.

Table \ref{tab:active_learning_exp} shows the conducted experiments to validate whether sampled tweets with this approach contribute to the model's ability to learn and generalize. In this experimental setting, the initial annotated batch of English tweets is split to 8000 training and 2000 testing tweets. A simple CNN model is trained and the macro average F1 score is calculated for the test set. This process is repeated 3 times using different additional annotated sets to train the classifier. The first set consists of 2000 randomly sampled tweets, the second set consists of 2000 tweets sampled from the [0.2-1.0] hate speech probability interval, and the third contains 2000 tweets sampled using the active learning approach. As we can see in Table \ref{tab:active_learning_exp}, random sampling does not improve the evaluation results at all. On the other hand, both hate probability and active learning approaches improved the macro-f1 evaluation with the latter being the best approach. In this experiment, it is obvious that active learning improves the learning process of the classifier by choosing the most appropriate tweets for annotation.

\subsection{Datasets}
\label{Datasets}

\begin{table}[ht]
    %\resizebox{\columnwidth}{!}{
    \centering
    \begin{tabular}{|c|c||c|c|}
    
        \hline 
        \multirow{2}{*}{\textbf{Language}}& \multirow{2}{*}{\textbf{Set}} & \multicolumn{2}{c|}{\textbf{Hate speech}}  \tabularnewline
        \cline{3-4} 
        & & \textbf{positive} & \textbf{negative}\tabularnewline
        \hline
        \hline 
        \multirow{3}{*}{EN} & train & 5804 & 68051\tabularnewline
        \cline{2-4}
        & test & 1355 & 16812 \tabularnewline
        \cline{2-4}
        & total & 7159 & 84863 \tabularnewline
        \hline 
        
        \multirow{3}{*}{DE} & train & 1361 & 33626\tabularnewline
        \cline{2-4}
        & test & 340 & 8707 \tabularnewline
        \cline{2-4}
        & total & 1702 & 42033\tabularnewline
        \hline 

        \multirow{3}{*}{ES} & train & 795 & 29355\tabularnewline
        \cline{2-4}
        & test & 199 & 7339 \tabularnewline
        \cline{2-4}
        & total & 994 & 36694\tabularnewline
        \hline 
        
        \multirow{3}{*}{FR} & train & 2163 & 29124\tabularnewline
        \cline{2-4}
        & test & 541 & 7281\tabularnewline
        \cline{2-4}
        & total & 2704 & 26405\tabularnewline
        \hline 
        
        \multirow{3}{*}{GR} & train & 913 & 48271\tabularnewline
        \cline{2-4}
        & test & 228 & 12069\tabularnewline
        \cline{2-4}
        & total & 1141 & 60340\tabularnewline
        \hline 
        
    \end{tabular}
    %}
    
    \caption{\label{tab:dataset_statistics} Number of positive and negative annotated tweets in different sets and languages available in our publicly available datasets}
\end{table}

The final datasets for English\footnote{https://zenodo.org/record/3520152\#.XcL0OnUzY5k}, German\footnote{https://zenodo.org/record/3520148\#.XcL04XUzY5k}, Spanish\footnote{https://zenodo.org/record/3520150\#.XcL1C3UzY5k}, French\footnote{https://zenodo.org/record/3520156\#.XcL1GHUzY5k} and Greek\footnote{https://zenodo.org/record/3520157\#.XcL1G3UzY5k} languages are hosted on Zenodo platform and are available after request. Each dataset contains tweets ids and their corresponding binary annotations. Table \ref{tab:dataset_statistics} shows the individual dataset statistics for each language. The datasets are provided in train/test sets, preserving the proportion of negative and positive samples.

% MAYBE: say why numbers are different

\section{Experimental study}
\label{Experimental study}

In this section we present the experimental pipeline that is followed in order to train and evaluate the hate speech detection models using the datasets described in the previous Section.
\subsection{Tweet pre-processing}
\label{Tweet pre-processing}

Due to the nature of Twitter data, there is a lot of noise
among words. Posted links or mentions do not provide
any useful information and need to be normalized. To achieve this, a state of the art tweet
normalization tool \cite{baziotis-pelekis-doulkeridis:2017:SemEval2}
is used, to tokenize and transform each tweet into a sequence of words. The
process involves Twitter handles normalization (e.g. @random\_user
becomes $<$ user $>$), emoji transformation (e.g. :( becomes $<$ sad $>$), lower
casing, as well as, URL, email and number removal. Furthermore, only basic punctuation is retained (e.g. .,?;'').

\subsection{Models}
\label{Models}

Following the latest trend in the literature, which shifts towards
the adoption of deep learning based methods, some of the latest
state of the art models for text classification are used. Deep learning models perform better than traditional methods in most NLP tasks, including hate speech detection tasks \cite{gamback_using_2017,park_one-step_2017}.
Furthermore,  an ensemble learning architecture is proposed, since it combines the predictive power of each individual classifier. Below we describe the deep learning architectures that are evaluated in this work.
\begin{description}
\item [{CNN.}] A simple Convolutional Neural Network model described in \cite{zhang_hate_2018}
acts as n-gram feature extractor. Using windows sizes of 2,3 and 4
this CNN model can extract bi-gram, tri-gram and quad-gram features.
The output of each CNN is then further down-sampled by a 1D max pooling
layer with a pool size of 4 and a stride of 4 for further feature
selection. After the concatenation of pooling layers, another 1D max
pooling layer is added and the output is fed to the final fully connected
layer.
\item [{Skipped}] \textbf{CNN (sCNN). }Extending the base CNN model in
order to capture features of words that are not next to each other,
Zhang et al. \cite{zhang_hate_2018} proposed Skipped CNN layer. Skipped
CNN applies a mask to a kernel window, skipping intermediate words
and associating words that are not directly near. According to the
authors, skipped CNNs can be considered as extractors of ‘skip-gram’
like features.
\item [{CNN}] \textbf{+ GRU}. Work in \cite{zhang_detecting_2018} added
a GRU layer followed by a global max pooling layer on top of CNN model.
The GRU layer captures sequence feature relations and learns to identify
dependencies between n-gram features
\item [{LSTM.}] A bidirectional LSTM model is created. After the embedding
layer, spatial dropout is introduced, which randomly masks 20\% of the
input words. To process the sequence of word embeddings, an
LSTM layer is used with 128 units. Next, a global max pooling
and an average max pooling layer are concatenated, flattening the output space by taking the highest and the average value in each timestep dimension,
respectively. The produced feature vector is  fed into the
final fully connected layer.
\item [{LSTM}] \textbf{+ Attention (aLSTM).} Attention mechanism is used
with success in many NLP tasks like in \cite{luong_effective_2015}.
Intuitively, attention is a mechanism that learns to favor features
that are more relevant to the classification task, by assigning weights to them.
This means that features that are not important to the task are multiplied
by smaller weights, while predictive features are multiplied by higher
weights. The attention layer is implemented based on \cite{zhou_attention-based_2016}
and it is applied to the LSTM model. Instead of taking the max and
average features in each timestep, an attention layer with 100
units is used to extract the important features of the LSTM layer. The output of the attention layer is then fed to the output layer.
\item [{Ensemble}] \textbf{(E)}. Aken et al. \cite{aken_challenges_2018}
proposed an ensemble model based on the assumption that classification
methods vary in their predictive power and may conduct specific errors.
The ensemble model in \cite{aken_challenges_2018} is trained with gradient
boosting decision trees. We used a simple dense neural network
ensemble architecture, forwarded the output predictions of the models
as inputs to a dense layer with 20 neurons, applied a dropout layer
with a ratio of 0.2 and finally the outputs features are forwarded to the
final output layer. Intuitively, this small neural network learns
to apply a weighted average based on the prediction probability of
each individual classifier.
\end{description}

\subsection{Implementation details}
\label{Implementation details}

For all methods discussed in this work, we use Keras \cite{chollet2015keras}
with Tensorflow \cite{tensorflow2015-whitepaper} backend and the
scikit-learn \cite{scikit-learn} library. Each model is trained for
10 epochs and a mini-batch of 64 tweets is used. Keras requires static
input sequences, meaning that the max number of words in a tweet has
to be predefined. Thus, the max sequence of words for
a tweet is set to 50, since, after experimentation, it is found that it does not affect performance.
Zero padding is used for sentences with less than 50 words. The first
layer for every model is an embedding layer. We initialize the embedding layer using pre-trained
word vectors for each language.
After conducting some preliminary experiments, the best pre-trained embedding choice for Greek and French language is using fastText embeddings \cite{grave2018learning}, trained on Common Crawl and Wikipedia. For English, Spanish and German language Glove embeddings \cite{pennington2014glove} achieve better evaluation results. Word2vec \cite{mikolov2013distributed} pre-trained embeddings are also tested. Note that the evaluation results among different embedding approaches do not exhibit significant differences.
Word vectors that do not exist in the pre-trained embeddings are randomly initialized, and the embedding layer is further fine-tuned during the training process.
To represent a padding token, zero initialization is used. For every model, the default parameters are used, as they are provided by the corresponding authors, unless stated otherwise. The l2 regularization
parameter is set to be $1e^{-3}$ for every layer. We treat hate speech detection as a binary classification problem.
The final fully connected layer is a sigmoid activation and outputs the hate speech probability. Binary
cross-entropy loss function and the Adam optimizer are used to train
the models.

\subsection{Evaluation setup}
\label{Evaluation setup}

In related literature, evaluation of the performance of hate speech
detection typically adopts the classic Precision, Recall and F1 metrics.
Precision measures the percentage of true positives among the predicted
hate speech tweets. Recall measures the percentage of true positives
among the ground truth hate speech tweets, and F1 calculates the harmonic
average of the two. The three metrics are applied to each dataset
class and an aggregated result is computed, either using micro-average
or macro-average. The first approach sums up the individual true positives,
false positives, and false negatives identified by a model, not taking
into consideration different classes to calculate overall Precision,
Recall and F1 scores. The second approach takes the average of the
Precision, Recall and F1 on different classes. Existing studies on
hate speech detection have primarily reported their results using
micro-average Precision, Recall and F1 \cite{badjatiya_deep_2017,gamback_using_2017,park_one-step_2017,waseem_are_2016,zhang_detecting_2018}.

As stated in \cite{zhang_hate_2018} and is made obvious in our dataset
statistics shown in Table \ref{tab:dataset_statistics}, a usual observation
in hate speech datasets is their highly imbalanced nature. In imbalanced
datasets, like the ones discussed in this paper, micro-averaging can inherently
hide the real performance of minority classes. Thus, a significantly
lower or higher F1 score on a minority class is unlikely to cause
a significant change in micro-F1 on the entire dataset. In a practical
application like hate speech detection, reporting micro-F1 on the entire
dataset will not properly reflect a model's performance on hateful
content as opposed to non-hate. Motivated by these observations, we
use the standard Precision (P), Recall (R) and F1 measures for evaluation
and report their macro averages(m-P, m-R, m-F1). Additionally, we provide F1 obtained on hate speech class (h-F1).

To train and evaluate the models for hate speech detection,
we use the training and test sets reported in Table \ref{tab:dataset_statistics},
respectively.

\subsection{Results}
\label{Results}

Table \ref{tab:hate_results} shows the evaluation results for the hate speech
class in each language. A first observation that highlights the imbalance between classes
is that F1 score for the hate class is  significantly lower compared to the macro F1 scores. This is expected because the number of negative annotated tweets
in the test dataset is significantly larger than positive annotated ones, as displayed
in Table \ref{tab:dataset_statistics}.

\begin{table}[ht]
    %\resizebox{\columnwidth}{!}{
    \centering
    \begin{tabular}{|c|c||c|c|P{1.1cm}|c|c|c|}
    
        \hline 
        \textbf{} & \textbf{metric} & \textbf{CNN} & \textbf{sCNN} & \textbf{CNN + GRU} & \textbf{LSTM} & \textbf{aLSTM} & \textbf{E}\tabularnewline
        \hline 
        \hline 
        \multirow{4}{*}{EN} & m-P & 0.81 & \textbf{0.83} & 0.80 & 0.77 & 0.79 & 0.80\\
        \cline{2-8} 
         & m-R & 0.78 & 0.78 & 0.80 & 0.78 & 0.79 & \textbf{0.82}\\
        \cline{2-8} 
         & m-F1 & 0.79 & 0.80 & 0.80 & 0.77 & 0.79 & \textbf{0.81}\\
        \cline{2-8}
         & h-F1 & 0.61 & 0.64 & 0.63 & 0.58 & 0.61 & \textbf{0.65}\\
        \cline{1-8}
        \cline{1-8}
        
        \multirow{4}{*}{DE} & m-P & 0.64 & \textbf{0.67} & 0.68 & 0.65 & \textbf{0.67} & \textbf{0.67}\\
        \cline{2-8} 
         & m-R & 0.67 & \textbf{0.71} & 0.68 & 0.65 & 0.66 & \textbf{0.71}\\
        \cline{2-8} 
         & m-F1 & 0.65 & \textbf{0.69} & 0.68 & 0.65 & 0.66 & \textbf{0.69}\\
        \cline{2-8}
         & h-F1 & 0.34 & \textbf{0.40} & 0.38 & 0.32 & 0.35 & \textbf{0.40}\\
        \cline{1-8}
        \cline{1-8}

        \multirow{4}{*}{ES} & m-P & 0.69 & 0.69 & 0.70 & \textbf{0.74} & 0.68 & 0.70\\
        \cline{2-8} 
         & m-R & 0.71 & \textbf{0.75} & 0.72 & 0.68 & 0.68 & 0.73\\
        \cline{2-8} 
         & m-F1 & 0.70 & \textbf{0.72} & 0.71 & 0.70 & 0.68 & \textbf{0.72}\\
        \cline{2-8}
         & h-F1 & 0.42 & \textbf{0.45} & 0.44 & 0.42 & 0.38 & 0.44\\
        \cline{1-8}
        \cline{1-8}
        
        \multirow{4}{*}{FR} & m-P & 0.81 & 0.81 & 0.83 & 0.80 & 0.80 & \textbf{0.84}\\
        \cline{2-8} 
         & m-R & 0.81 & \textbf{0.82} & 0.81 & 0.77 & \textbf{0.82} & 0.81\\
        \cline{2-8} 
         & m-F1 & 0.81 & 0.82 & 0.82 & 0.78 & 0.81 & \textbf{0.83}\\
        \cline{2-8}
         & h-F1 & 0.65 & 0.66 & 0.66 & 0.64 & 0.64 & \textbf{0.67}\\
        \cline{1-8}
        \cline{1-8}
        
        \multirow{4}{*}{GR} & m-P & 0.81 & \textbf{0.87} & \textbf{0.87} & 0.86 & 0.86 & \textbf{0.87}\\
        \cline{2-8} 
         & m-R & \textbf{0.78} & 0.77 & 0.75 & 0.75 & 0.75 & \textbf{0.78}\\
        \cline{2-8} 
         & m-F1 & 0.79 & 0.81 & 0.80 & 0.80 & 0.80 & \textbf{0.82}\\
        \cline{2-8}
         & h-F1 & 0.59 & 0.63 & 0.60 & 0.60 & 0.60 & \textbf{0.65}\\
        \cline{1-8}
        
    \end{tabular}
    %}
    
    \caption{\label{tab:hate_results}The evaluation results for hate speech class}
\end{table}

By inspecting each language separately, we notice that there are no significant performance differences between all models in terms of macro F1. However, in terms of individual models,
the sCNN model seems to generally exhibit the best performance. Some exceptions are observed, as in the case of the Spanish language, where the  LSTM model performs better in terms of macro Precision, and in the Greek language, where the CNN model has better macro Recall evaluation.
sCNN seems to be the most compelling feature extractor for hate speech as it achieves the best F1 score for the hate class, among individual models. This also corresponds to overall better macro F1 by sCNN compared to other methods. For the case of the French language, CNN+GRU model performs on par with sCNN model.

Additionally, the combination of all individual models in the ensemble
model (E) yielded even better results in terms of macro F1. The ensemble model
had the best macro F1 score, as it manages to perform well both in
terms of macro Precision and macro Recall. The ensemble model also exhibits the best performance in terms of hate speech F1 score. The only exception is observed in the Spanish language, where sCNN model scores a higher F1 score for the hate speech class.
 
Another observation is that the evaluation for English, French and Greek, specifically in terms of F1 score in the positive class, is significantly better when compared with the Spanish and the German languages. This is potentially due to the fact that there are less positive samples in the related datasets. Our goal is to continue expanding the datasets and specifically address the issue for these two languages.

\section{Conclusion}
\label{Conclusion}

In this work, we try to tackle hate speech directed at journalists on social media. 
%We study hate speech in five different languages. 
To accomplish this, we define hate speech in a way that takes into consideration the journalistic point of view, and is simple enough to be used by non specialists. Using this definition we create labeled datasets in five different languages. During data annotaion, a comprehensive annotation strategy is followed. The generated datasets are made publicly available to assist further research efforts. Furthermore, we use these datasets to train various state-of-the-art deep learning architectures, while at the same time, we propose an ensemble model that outperforms all individual models.

Another major contribution of this study is its annotation pipeline. To increase the number of positive annotations, we employed keyword-based sampling and also an approach that uses available datasets from other related works. Namely, the dataset-based sampling approach trains detection models using related datasets and samples tweets based on the output probability of the model. This approach, is similar to transferring the definition of hate speech from other works and use it to sample tweets. This is particularly useful in cases where the definition of hate speech in these datasets has a wider scope than the proposed definition. It is obvious that these approaches introduce bias to the resulting datasets. Although bias is inevitable, we propose ways to mitigate it. Lowering the sampling threshold of the detection models, combining multiple datasets and adding randomly sampled data, are good approaches to deal with bias. 

Building on the initial annotation stage, we also presented an active learning approach. The contribution of this stage is that, besides the high percent of positive annotations in the resulting batches, these batches contain data that can contribute to the learning process of the models. Specifically, data sampled with this approach generally deviate from the already labeled data and consequently improve the generalization of the model.

We believe that the presented annotation process can be applied to other domains, beyond the scope of this work. Specifically, the two sampling stages can be very effective approaches when annotating a large corpus of data. The only input that is required for these approaches is one or more relevant datasets or a curated list of keywords.

%TO DO: say jounalists will add more
As future steps, we plan to keep expanding the datasets with new tweets. To this end, we have developed an alert monitoring mechanism for journalists that supports further annotation of tweets. Using these tweets, we plan to retrain our models in frequent intervals.
Additionally, we aim to investigate new active learning techniques in order to
choose more informative tweets that improve the models' ability to learn and generalize.
Another issue we will focus on, is the imbalance between the positive
and negative classes. To alleviate this, we will explore ways to
fetch more hateful content in a more unbiased manner. Finally, we will experiment with
some state of the art deep learning architectures for natural language
processing like BERT \cite{devlin2018bert} or ULMFiT \cite{howard2018universal}.

\section{Acknowledgements}
\label{Acknowledgements}

This work was supported by the Rights, Equality and Citizenship programme of the European Union (2014-2020) under grant agreement number 785679. 

% \section*{\refname}
\bibliographystyle{elsarticle-num}
%\biboptions{numbers,sort&compress}
\bibliography{references_final.bib}

\end{document}